\documentclass[10pt,conference]{IEEEtran}
\IEEEoverridecommandlockouts
\usepackage{url}
\usepackage{amsmath,amssymb,amsfonts}
\usepackage[linesnumbered,ruled,vlined]{algorithm2e}
\usepackage{graphicx}
\usepackage{algpseudocode}
\usepackage{physics}
\usepackage{textcomp}
\usepackage{stfloats}
\usepackage[colorinlistoftodos]{todonotes}
\usepackage{hyperref}
\usepackage{cite}
\usepackage{xcolor} 
\usepackage{booktabs}
\usepackage{amsmath} % 用于数学公式
\usepackage{arydshln}
\usepackage{soul}
\usepackage{svg}
\usepackage{lipsum,mwe,cuted}
\def\BibTeX{{\rm B\kern-.05em{\sc i\kern-.025em b}\kern-.08em
    T\kern-.1667em\lower.7ex\hbox{E}\kern-.125emX}}
\usepackage{caption}
\begin{document}

% \title{Backbone-Guided Integration of QAOA with Tabu Search for QUBO Model Optimization
% }
% \title{Backbone-Adaptive QAOA: Variable Fixation Strategies for Scalable Quantum Optimization on Resource-Constrained Hardware}
\title{Hierarchical Quantum Optimization via Backbone-Driven Problem Decomposition: Integrating Tabu-Search with QAOA}
% \title{Backbone Variable Fixation in QAOA: Enhancing NISQ-Era Optimization via Hybrid Tabu-Quantum Strategies}
%\\

% \author{\IEEEauthorblockN{Minhui Gou}
% \IEEEauthorblockA{\textit{Computer and Information Sciences} \\
% \textit{Computer Technology} \\
% \textit{China University of Petroleum}\\
% Beijing, China \\
% }

% }

\author{
    \IEEEauthorblockN{
    Minhui Gou \IEEEauthorrefmark{1}\IEEEauthorrefmark{4}, 
    Zeyang Li\IEEEauthorrefmark{1}\IEEEauthorrefmark{4}, 
    Hong-Ze Xu \IEEEauthorrefmark{1} \IEEEauthorrefmark{2},
    Changbin Lu \IEEEauthorrefmark{7},
    Jing-Bo Wang \IEEEauthorrefmark{1} \IEEEauthorrefmark{2}\\
    Yukun Wang\IEEEauthorrefmark{4}\IEEEauthorrefmark{5} \thanks{wykun@cup.edu.cn},
    Meng-Jun Hu\IEEEauthorrefmark{1} \IEEEauthorrefmark{2} 
            \thanks{humj@baqis.ac.cn},
    Dong E Liu \IEEEauthorrefmark{1}\IEEEauthorrefmark{2} 
               \IEEEauthorrefmark{3} \IEEEauthorrefmark{6}
               \IEEEauthorrefmark{8}
               \thanks{dongeliu@mail.tsinghua.edu.cn},
    Wei-Feng Zhuang\IEEEauthorrefmark{1} 
                    \IEEEauthorrefmark{2} 
    \thanks{zhuangweifeng@baqis.ac.cn}
    }
    \IEEEauthorblockA{
        \IEEEauthorrefmark{1}Beijing Academy of Quantum Information Sciences, Beijing 100193, China
        }
    \IEEEauthorblockA{
        \IEEEauthorrefmark{2} Beijing Key Laboratory of Fault-Tolerant Quantum Computing, Beijing 100193, China
    }
    \IEEEauthorblockA{
        \IEEEauthorrefmark{3} State Key Laboratory of Low Dimensional Quantum Physics,  \\ Department of Physics, Tsinghua University, Beijing, 100084, China
    }
    \IEEEauthorblockA{
        \IEEEauthorrefmark{4} College of Artificial Intelligence, China University of Petroleum, Beijing 11414, China
        }
     \IEEEauthorblockA{
        \IEEEauthorrefmark{5} State Key Laboratory of Processors, Institute of Computing Technology, CAS, Beijing, 100190, China
    }
    \IEEEauthorblockA{
        \IEEEauthorrefmark{6} Frontier Science Center for Quantum Information, Beijing 100184, China
    }
    \IEEEauthorblockA{
        \IEEEauthorrefmark{7} School of Computer Science and Technology, Anhui University of Technology, Maanshan 243002, China
    }
    \IEEEauthorblockA{
        \IEEEauthorrefmark{8} Hefei National Laboratory, Hefei 230088, China\\
    }
}

\maketitle

\begin{abstract}
\indent As quantum computing advances, quantum approximate optimization algorithms (QAOA) have shown promise in addressing combinatorial optimization problems. However, the limitations of Noisy Intermediate Scale Quantum (NISQ) devices hinder the scalability of QAOA for large-scale optimization tasks. 
To overcome these challenges, we propose Backbone-Driven QAOA, a hybrid framework that leverages adaptive Tabu search for classical preprocessing to decompose large-scale quadratic unconstrained binary (QUBO) problems into NISQ-compatible subproblems. In our approach, adaptive Tabu search dynamically identifies and fixes backbone variables to construct reduced-dimensional subspaces that preserve the critical optimization landscape. These quantum-tractable subproblems are then solved via QAOA, with the resulting solutions iteratively refining the backbone selection in a closed-loop quantum-classical cycle.
Experimental results demonstrate that our approach not only competes with, and in some cases surpasses, traditional classical algorithms but also performs comparably with recently proposed hybrid classical-quantum algorithms. 
Our proposed framework effectively orchestrates the allocation of quantum and classical resources, thereby enabling the solution of large-scale combinatorial optimization problems on current NISQ hardware.

\end{abstract}

% \begin{IEEEkeywords}
% Quantum Optimization, QAOA, Backbone Variables, 
% Hybrid Quantum-Classical Algorithms
% \end{IEEEkeywords}

\section{Introduction}
Combinatorial optimization problems play a crucial role in fields such as finance \cite{markowitz1952portfolio}, healthcare \cite{abdalkareem2021healthcare}, bioinformatics \cite{rbioinformatics}, and production scheduling \cite{r51}, yet their NP-hard nature and vast solution spaces pose significant challenges for efficient solution methods
\cite{np,hartmanis1982computers}.
Over the decades, heuristic and metaheuristic algorithms—such as ant colony optimization \cite{ACO}, genetic algorithms \cite{GA}, and Tabu search \cite{r49}—have been successfully employed to find satisfactory solutions within reasonable time frames. However, as the problem scale increases, these classical methods face exponential growth in computational cost.

Quantum computing provides a powerful alternative by harnessing intrinsic quantum resources—such as superposition and entanglement—to encode and simultaneously probe an exponentially large space of potential solutions
\cite{shor1994algorithms,arute2019quantum,googlewillow,chitambar2019quantum,r54}.
In the era of Noisy Intermediate-Scale Quantum (NISQ) devices, the limited number of qubits and the presence of noise remain key obstacles to realizing practical quantum advantage \cite{brooks2019beyond}. Nonetheless, these constraints have inspired the development of Variational Quantum Algorithms (VQAs), which are designed to work within the capabilities of near-term quantum hardware \cite{r11, peruzzo2014variational,biamonte2017quantum}. Among them, the Quantum Approximate Optimization Algorithm (QAOA) \cite{r11} has gained significant attention as a leading candidate for solving combinatorial optimization problems. Other notable VQAs include the Variational Quantum Eigensolver (VQE) \cite{peruzzo2014variational}, and frameworks for quantum machine learning \cite{biamonte2017quantum}.
QAOA encodes combinatorial problems as QUBO instances, maps them to Ising Hamiltonians, and optimizes the solution using a parameterized quantum circuit coupled with classical feedback. While increasing the circuit depth ($p$) generally improves QAOA’s accuracy \cite{r11}, scalability remains a challenge due to noise, limited qubits, and the vast parameter space leading to barren plateaus. To address these issues, various strategies have been explored, including effective parameter initialization, adaptive mixer selection, and hybrid classical-quantum decompositions \cite{warm-start, gnn-qaoa, qa-qaoa, alignment, r25, greedy-qaoa, depth-qaoa}. Notably, decomposition techniques partition large problems into smaller subproblems that fit on NISQ devices \cite{r31, r32}.

In our work, we propose a novel hybrid approach that extends the partitioning-based optimization strategy initially developed by D-Wave, which combines Quantum Annealing (QA) with Tabu Search \cite{boost2017partitioning}. Our method incorporates backbone variable identification via Tabu Search into the QAOA framework, constructing a hierarchical structure over the QUBO problem. By selecting key backbone variables, the original problem is decomposed into manageable subproblems that can be efficiently solved using QAOA. To evaluate our framework, we execute the QAOA on a superconducting quantum processor accessed through the \textit{Quafu} quantum cloud platform developed by the Beijing Academy of Quantum Information Sciences (BAQIS). Experimental results demonstrate the effectiveness and efficiency of our approach. The proposed framework can dynamically balance classical and quantum resources, thereby enhancing scalability and performance for large-scale combinatorial optimization on NISQ hardware.
% In this work, we generalized partitioning optimization strategy that develop by D-wave
% which combine the Annealing (QA) algorithm with Tabu search, to propose a novel hybrid approach based on backbone variable identification via Tabu search, which constructs a hierarchical structure for the QUBO problem. By selecting key “backbone” variables, the problem is decomposed into manageable subproblems that are solved using QAOA, and the solutions are then integrated back into the global optimization process. Our framework dynamically balances classical and quantum resources, thereby enhancing scalability and performance on large-scale combinatorial optimization tasks.

% \section{Preliminaries and Notations}
\section{Background}
% To set up our notation and terminology we start out with a brief review
% of both QUBO, Tabu search and QAOA.

\subsection{Quadratic Unconstrained Binary Optimization.}
The QUBO framework \cite{r2} represents a widely used mathematical formulation for encoding and solving a broad class of NP-hard combinatorial optimization problems \cite{QUBO}. Within the QUBO framework, problems are formulated as optimization tasks involving binary variables, to minimize a target function composed of both linear and quadratic terms. 
The objective function of a QUBO problem is commonly represented as
\begin{equation} 
f(\boldsymbol{x}) = \sum_{i} Q_{ii} x_i + \sum_{i<j} Q_{ij} x_i x_j, \label{eq:quadratic} \end{equation}  
where $Q_{ii}$ corresponds to the coefficients of the linear terms, $Q_{ij}$ captures the pairwise interaction terms, and $\boldsymbol{x} = (x_1, x_2, \dots, x_n)$ denotes a vector of binary variables with $x_i \in \{0, 1\}$.The objective of the problem is to find a set of values for $x_i$ that minimizes or maximizes the objective function $f(\boldsymbol{x})$.

QUBO problems are particularly significant in quantum computing because they can be mutually converted with the Ising model, thereby transforming into Hamiltonians suitable for solving on quantum systems\cite{r19}. This transformation enables the possibility of using quantum computers to solve NP-hard problems, especially through methods like quantum annealing (e.g., Quantum Annealing in D-Wave systems)\cite{d-wave} and quantum variational optimization algorithms (such as QAOA)\cite{r20}. Hence, QUBO provides a new pathway for solving combinatorial optimization problems, combining the characteristics of quantum computing, and potentially holding immense promise for addressing certain complex problems in the future.

The transformation from a QUBO formulation to an Ising Hamiltonian entails converting the objective function of the QUBO problem into an equivalent Hamiltonian in the Ising model framework. This process involves mapping the binary variables and weights of the QUBO formulation to spin eigenvalue variables and coupling coefficients in the Ising model. The relationship between the QUBO binary variable $x_i \in \{0, 1\}$ and the Ising spin eigenvalue $z_i \in \{-1, +1\}$ is given by
\begin{equation}
\begin{aligned}
x_i = \frac{1 - z_i}{2},
\end{aligned}
\label{eq:x_i_to_sigmaz}
\end{equation}
Under this mapping, the QUBO Hamiltonian can be expressed in the standard Ising form
\begin{equation}
H = \sum_{i<j} J_{ij} Z_i Z_j + \sum_i h_i Z_i + C,
\label{eq:Ising}
\end{equation}
where Pauli-$Z_i$ matrix's eigenvalue is $z_i$,
$J_{ij} = Q_{ij}/4$ represents the coupling strength between spins $i$ and $j$, and $h_i = -\big(Q_{ii}/2 + \sum_{j\neq i} Q_{ij}/4\big)$ corresponds to the local effective field acting on the $i$-th spin. The constant offset $C = \frac{1}{4}\sum_{i \neq j} Q_{ij} + \frac{1}{2}\sum_i Q_{ii}$ accounts for the overall energy shift and is often omitted in optimization problems as it does not influence the spin configuration that minimizes the Hamiltonian. The Ising Hamiltonian provides a spin-based representation of the QUBO problem, preserving the equivalence of their solution spaces. 

\subsection{Tabu-Search}
The Tabu-Search (TS) algorithm \cite{glover1986future} is a well-established metaheuristic for combinatorial optimization. Its key strength lies in escaping local optima by temporarily accepting inferior solutions, guided by adaptive memory structures such as the Tabu List. While TS does not guarantee convergence to the global optimum, it often achieves high-quality solutions in practice \cite{glover1997tabu,r2005tabu}. 
% For QUBO case, we adopt the A
This process relies on the following  Two key components\cite{rtabututorial}:
\begin{itemize} \item \textbf{Neighborhood Search:} Tabu-search is based on neighborhood search, in which a series of new solutions, known as neighborhood solutions, are generated by making small modifications to the current solution. Let $S$ represent the current solution, and let $N(S)$ denote the neighborhood of $S$. The neighborhood $N(S)$ is the set of solutions that can be obtained by applying certain small changes (e.g., flipping a variable or exchanging elements) to $S$: 
\begin{equation}
N(S) = \{ S' \mid S' = S + \Delta S \},  
\end{equation}
 where $\Delta S$ represents a small modification applied to the current solution $S$. The objective is to find a new solution $S' \in N(S)$ that improves the objective function value.

\item \textbf{Tabu-List:}
% A distinctive feature of Tabu search is the use of a Tabu-List to avoid repeated exploration of the visited solution space. 
The Tabu-List \( T \) stores elements that have been recently visited and marked as tabu, i.e., solutions that are not allowed to be revisited during the tabu period. The updated search process can be described as:
\[
S_{t+1} = \arg \max_{S' \in N(S_t) \setminus T} f(S'),
\]
where \( S_t \) is the current solution at iteration \( t \), \( f(S) \) is the objective function, and the neighborhood \( N(S_t) \setminus T \) excludes solutions that are in the Tabu-List. The Tabu period for objects stored in the Tabu-List is known as the Tabu-Term, denoted as \( \tau \), which ensures that solutions are not revisited for \( \tau \) iterations, thus promoting a more diverse exploration of the solution space.
\end{itemize}

We apply the above mechanisms as a preprocessing step for QUBO problems. The Tabu Search algorithm explores the neighborhood of the current solution to identify improved candidates, while the Tabu List and tabu tenure are used to guide the search away from previously visited (and potentially suboptimal) regions of the solution space. As the classical component of our hybrid framework, the precision of the tabu search can be flexibly adjusted to control the quality of the preprocessing solution. This provides a mechanism for dynamically managing classical resource usage within our framework. Further implementation details are provided in Section \ref{sub:BDSW-QAOA}.

\subsection{Quantum Approximate Optimization Algorithm}
QAOA is a hybrid algorithm that combines quantum and classical computing. It was originally proposed by Farhi et al\cite{r11}. Its alternating structure, comprising parameterized quantum operations and classical optimization, along with its suitability for shallow to medium-depth circuits, makes it suited for implementation on noisy intermediate-scale quantum (NISQ) devices. There are some experimental results showing that QAOA is usually not limited by small spectral gaps\cite{r25}, which makes it the best choice for solving complex combinatorial optimization problems in finite coherence time using NISQ devices. 

The core idea of QAOA is to encode the objective function of the optimization problem, like Eq.\ref{eq:quadratic},
as a cost Hamiltonian equation $\hat{H}_C$, which takes the Ising model form in Eq.\ref{eq:Ising}.
The eigenvalues of $\hat{H}_C$ correspond directly to the possible values of the objective function, with the ground state representing the optimal solution to the problem. To explore the solution space efficiently, QAOA utilizes a mixing Hamiltonian
$\hat H_M$, which is typically chosen to be a sum of single-qubit Pauli-$X$ operators $\hat{H}_M= \sum_{i=1}^N \hat X_i$. The quantum system is initialized in the ground state of $\hat{H}_M$, denoted as $|s\rangle$,
\begin{equation}
|s\rangle = |+\rangle ^{\otimes n } = 
\frac{1}{\sqrt{2^n}}\sum_{\mathbf{x}\in\{0,1\}^n}|\mathbf{x}\rangle.
\end{equation}
Starting from the initial state $|s\rangle$, the algorithm alternates between evolving the state under the cost Hamiltonian $\hat{H}_C$ and the mixing Hamiltonian $\hat{H}_M$. Each layer of the algorithm applies two unitary operations, $U_C(\gamma) = e^{-i\gamma \hat{H}_C}$ and $U_M(\beta) = e^{-i\beta \hat{H}_M}$, where $\gamma$ and $\beta$ are the variational parameters. After $p$ layers, the quantum state can be expressed as
\begin{equation}
    |\psi(\vec{\gamma}, \vec{\beta})\rangle = \prod_{k=1}^p U_M(\beta_k) U_C(\gamma_k) |s\rangle,
\end{equation}
where $\vec{\gamma} = (\gamma_1, \gamma_2, \dots, \gamma_p)$ and $\vec{\beta} = (\beta_1, \beta_2, \dots, \beta_p)$ are vectors of variational parameters to be optimized. 

\begin{figure}[htp]
    \centering
    \includegraphics[width=0.5\textwidth,height=0.25\textheight]{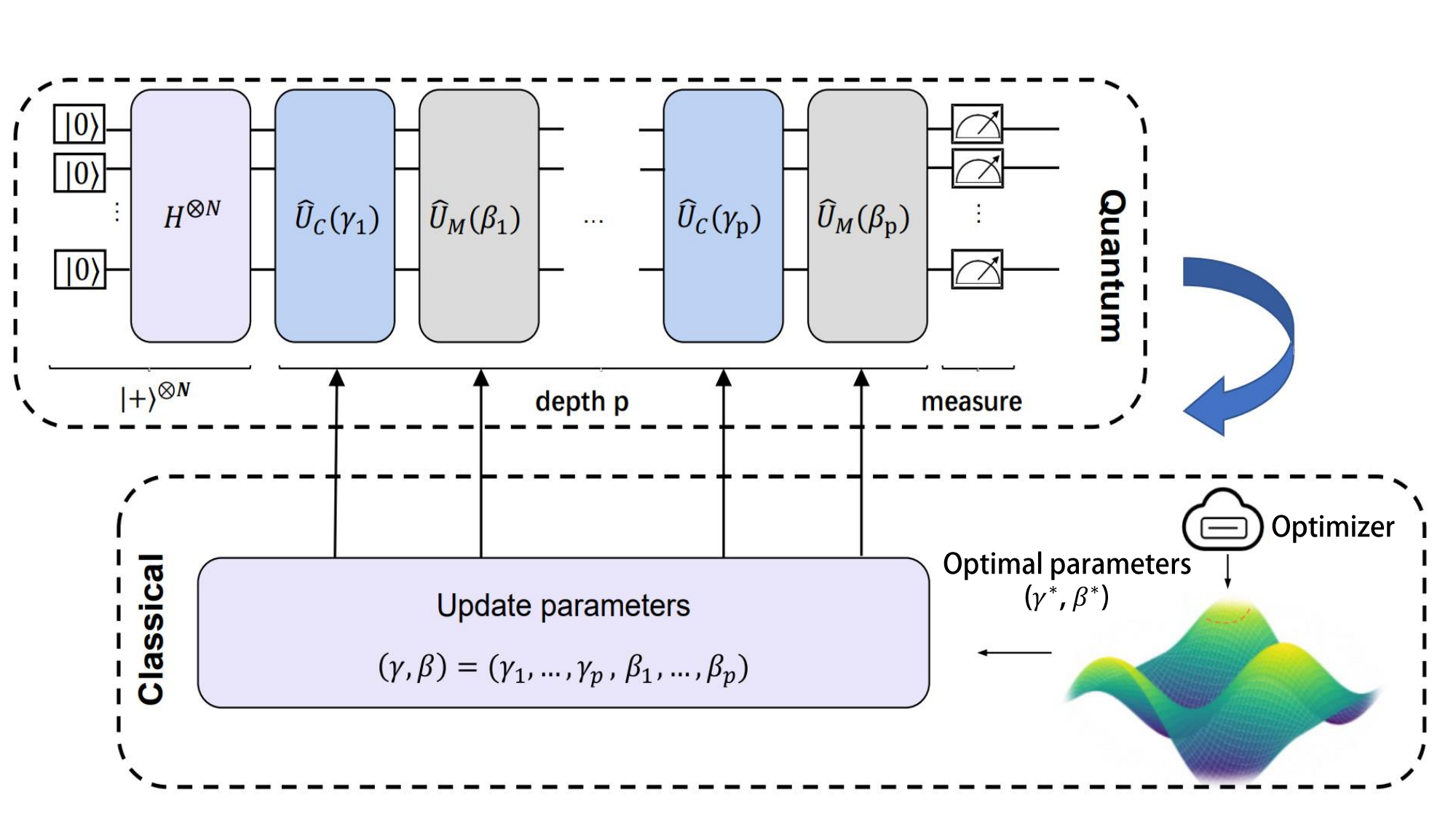}
    \caption{QAOA framework. QAOA as a type of VQA, which contains both classical and quantum parts. In the quantum part, a quantum system consisting of alternating p-level circuits $\hat{U}_C(\gamma)$ and $\hat{U}_M(\beta)$ acts on the maximal superposition state. For the classical part, it uses a classical optimizer to update the parameters of the circuit based on the bit strings obtained from measurements.}
    \label{fig:galaxy}
\end{figure}

The goal of QAOA is to determine the parameters $\vec{\gamma}$ and $\vec{\beta}$ that maximize the expectation value of the cost Hamiltonian
\begin{equation}
F_p(\vec{\gamma}, \vec{\beta}) = \langle \psi_p(\vec{\gamma}, \vec{\beta}) | \hat{H}_C | \psi_p(\vec{\gamma}, \vec{\beta}) \rangle,
\end{equation}
effectively finding the quantum state with the highest overlap with the ground state of $\hat{H}_C$. This is achieved by optimizing the variational parameters $(\vec{\gamma}, \vec{\beta})$ through classical algorithms to identify the optimal parameter set $(\vec{\gamma}^\ast, \vec{\beta}^\ast)$ that maximizes $F_p(\vec{\gamma}, \vec{\beta})$
\begin{equation}
(\vec{\gamma}^\ast, \vec{\beta}^\ast) = \arg\max_{\vec{\gamma}, \vec{\beta}} F_p(\vec{\gamma}, \vec{\beta}).
\end{equation}
The optimization is conducted classically, with the measured outcomes from the quantum circuit guiding iterative updates of the parameters. The workflow of QAOA, as described above, is illustrated in Fig.~\ref{fig:galaxy}, which outlines the alternation between quantum evolution and classical optimization.

An important metric for evaluating the performance of QAOA is the approximation ratio, denoted by $\alpha$. This ratio quantifies the algorithm's effectiveness to approximate the optimal solution of the classical optimization problem and is defined as
\begin{equation}
\alpha = \frac{F_p(\vec{\gamma}^\ast, \vec{\beta}^\ast)}{C_{\mathrm{max}}},
\end{equation}
where $F_p(\vec{\gamma}^\ast, \vec{\beta}^\ast)$ is the maximum value of the expectation of the cost Hamiltonian achieved by the algorithm, and $C_{\mathrm{max}} = \max_{\mathbf{x}} C(\mathbf{x})$ represents the true maximum of the classical cost function. The approximation ratio $\alpha$, bounded by $0 \leq \alpha \leq 1$, provides a normalized measure of how closely the quantum algorithm approaches the optimal classical solution. A higher value of $\alpha$ indicates better performance, making it a key benchmark for assessing QAOA’s effectiveness in solving combinatorial optimization problems.

As the number of layers $p$ in QAOA increases, the approximation ratio typically improves. Achieving specific approximation ratios often necessitates a sufficiently large number of layers \cite{blekos2024review,lotshaw2021empirical}. Furthermore, theoretical lower bounds exist on the number of QAOA rounds required to guarantee certain levels of approximation performance\cite{benchasattabuse2023lower}. In the numerical results presented in this study, we primarily compare the efficiency of our proposed algorithm to that of other state-of-the-art algorithms, using the approximation ratio as the key performance metric.

\begin{figure*}[!htb]
    \centering
    \includegraphics[width=1.0\textwidth]{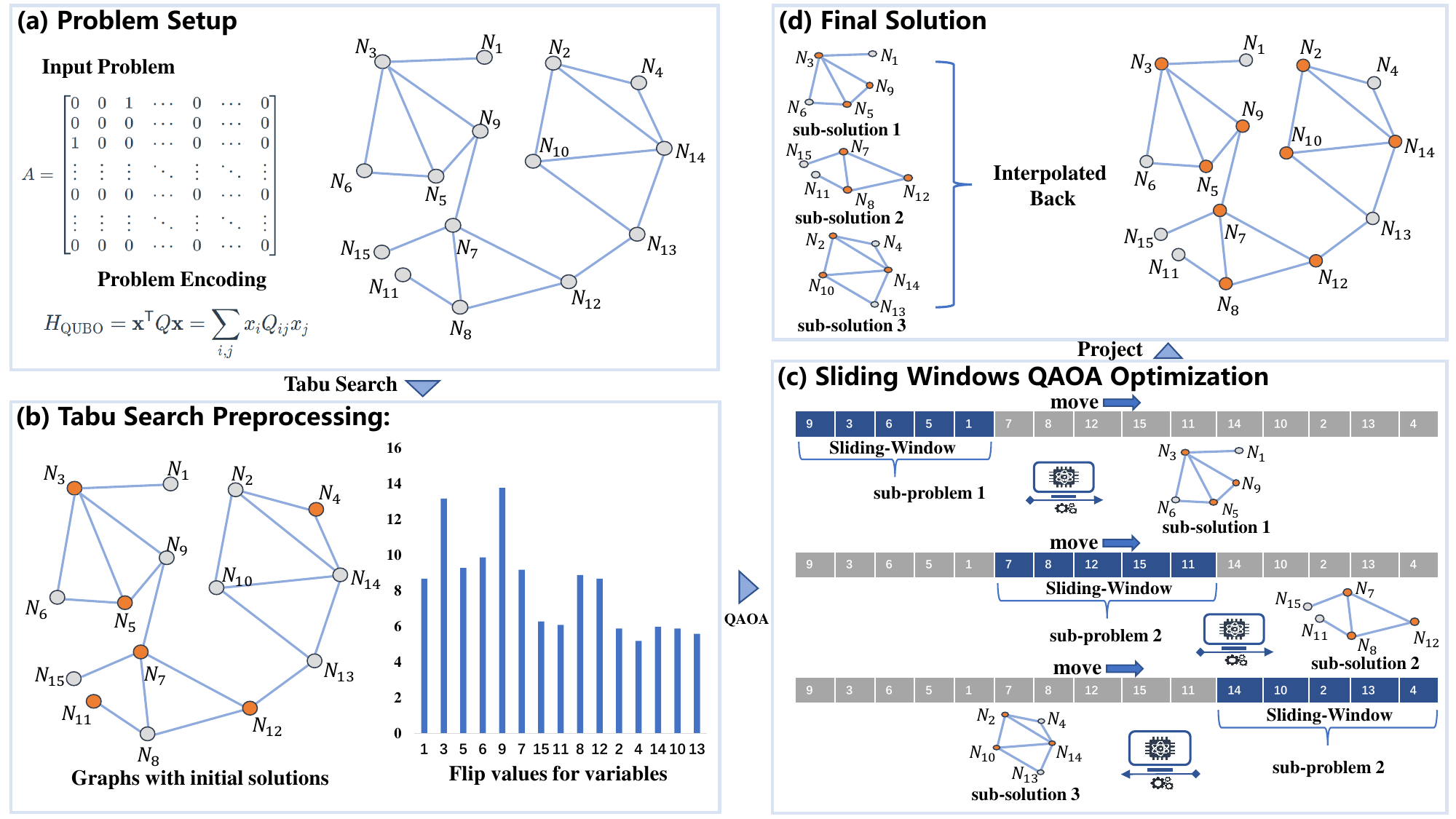}
    \caption{Overview of BDSW-QAOA. 
    (a) The algorithm begins by converting the input problem's adjacency matrix into a QUBO formulation and obtaining an initial solution for further processing.
    (b) In the Tabu search preprocessing phase, a preliminary solution is generated, and backbone variables are identified and ordered according to their flip cost.
    (c) A sliding window strategy is applied to the ordered backbone set to partition the problem into subproblems that match the NISQ device's capacity; these subproblems are then solved via QAOA.
    (d) Finally, the best-performing subproblem solution is projected back onto the original QUBO problem to reconstruct the full solution.}
    % (a) Firstly, we receive the adjacency matrix of the problem as input and then convert it to QUBO. 
    % (b) In the Tabu search preprocessing phase, We get an preprocessed solution by tabu search, and and backbone variables set the by flip cost ordering. (c) We utilize the sliding window to construct the subproblems to be solved based on the ordered backbone set, the subproblems' size is according to the size of the NISQ device, and subproblems are solve by QAOA. (d) After solving the subproblems in different windows in backbone variables set, we select the best solution to interploed back the the original qubo problem's solution to achieve the final solution.}
    \label{fig:overview}
\end{figure*}

\section{Backbone-Driven and Sliding Window QAOA}
\label{sub:BDSW-QAOA}
To improve the scalability of QAOA for large-scale combinatorial optimization problems, we employ backbone variable identification to decompose the original problems into reduced-scale subproblems that are more suitable for solving on current NISQ devices. The notion of backbone variable originates from satisfiability problems (SAT), where they are defined as a set of variables that maintain consistent values across all solutions of satisfiable instance \cite{r28,r29}.
In the context of combinatorial optimization, backbone variables are ideally those fixed in all optimal solutions. However, identifying backbone variables in practice requires knowledge of the global solution space. which is computationally infeasible for large-scale problems due to the exponential complexity involved.

We employ the concept of "strongly determined variables" \cite{glover2005adaptive} to approximate backbone variables during optimization. Given a candidate solution \( \mathbf{x}' = (x_1', x_2', \ldots, x_n') \), we quantify variable determinacy through a flip-cost metric derived from the quadratic objective function in Eq. \ref{eq:quadratic}. Specifically, for each variable \( x_i \), we calculate the cost difference \( \Delta C_i \) incurred by flipping its value from \( x_i' \) to \( 1 - x_i' \),

\begin{equation}
\Delta C_i = f'' - f' = (1 - 2x_i') \left( Q_{ii} + \sum_{j \in N, j \neq i} Q_{ij} x_j' \right),
\label{eq:flip_cost}
\end{equation}
where $f^\prime$ is the cost of the original solution, and $f^{\prime \prime}$ represents the cost after flipping $x_i^\prime$. Variables are ranked by descending flip cost \(|\Delta C_i|\), where larger magnitudes indicate stronger determination. 
 
After selecting the top-$k$ backbone variables ranked by $|\Delta C_i|$ to form the backbone set $\mathcal{B} = \{ x_{(1)}, \ldots, x_{(k)} \}$ (where $x_{(1)}$ maximizes $|\Delta C_i|$), we implement a step-by-step sliding window strategy compatible with NISQ constraints. To formalize the subproblem construction, consider a window $S \subseteq \mathcal{B}$ containing $n_{\text{w}}$ variables ($n_{\text{w}} \leq N_{\text{qubit}}$), where $N_{\text{qubit}}$ denotes the NISQ device's qubit capacity. Non-window variables $x_j \notin S$ are fixed to their Tabu-optimized values $x_j^*$, inducing a reduced QUBO over $S$ \cite{d-wave}:
\begin{equation}
f_S(\mathbf{x}_S) = \sum_{i \in S} \left( Q_{ii} + d_i \right)x_i + \sum_{\substack{i,j \in S \\ i < j}} Q_{ij}x_ix_j.
\label{eq:subqubo}
\end{equation}
The cross-coupling term \( d_i \), quantifying interactions between $ x_i $ and $ \bar{S} $, is given by
\begin{equation}
d_i = \sum\nolimits_{j \notin S} (Q_{ij} + Q_{ji}) x_j^\ast,
\label{eq:di}
\end{equation}
thereby preserving QUBO equivalence under variable fixation.
The implementation workflow is as follow:
\begin{itemize}
    \item Initial window: Initialize the first window as $\mathcal{W}_1 = (x_{(1)}, x_{(2)}, \ldots, x_{(n_{\text{w}})}) $, where $ x_{(i)} \in \mathcal{B} $ are the top-$ n_{\text{w}} $ ranked backbone variables $(n \le k)$.
    \item QAOA Optimization: 
    Apply QAOA to the window-restricted QUBO (Eq.~\ref{eq:subqubo}), optimizing $ \mathbf{x}_{\mathcal{W}_m}$ while enforcing $x_j^\ast$  -fixation for $j \notin \mathcal{W}_m$.
    \item Window shift: Slide the window to \( \mathcal{W}_{m+1} = (x_{(m+1)}, \ldots, x_{(m+n_{\text{w}})}) \), dynamically update \( d_i \) as defined in Eq.~\ref{eq:di} based on the latest \( x_j^* \), then re-apply the QAOA optimization (Step 2) for the new window.
\end{itemize}
This generates \( k - n + 1 \) overlapping subproblems where each backbone variable participates in \( n \) optimizations. At each iteration, the solution obtained from the subproblem (i.e., the optimized values of the $n$ variables in the current window) is incorporated into the original problem by updating the corresponding variables in the global solution. We then compute the objective function value of this updated global solution and compare it to the previous global solution. If the updated solution yields a better objective function value, we accept it as the new global solution. This ensures that the global solution progressively improves as the sliding window advances through the backbone variables.

Within our hybrid quantum-classical framework, the backbone variable driven and sliding-window QAOA (BDSW-QAOA) operate within the hierarchical structure of a Tabu-search metaheuristic framework. The whole algorithm of our approach is shown in Algorithm \ref{alg:bdsw_qaoa}. For the preprocessing Tabu search preprocessing stage, the Tabu tenure vector $\mathcal T  \in \mathbb N^n$ is initialized such that each entry $\mathcal T_i$ is set to zero, indicating that all variables are initially free to be flipped. At each iteration $t$, the algorithm computes the flip cost $|\Delta C_i|$ for all variables, determining how the solution would change if a specific variable $x_i$ were flipped. The variable with the highest flip cost $|\Delta C_i|$  is then selected for flipping, provided that its tabu tenure $\mathcal T_i = 0$, meaning it is not currently prohibited from being flipped. When a variable $x_j$ is flipped, its corresponding tabu tenure is updated to $\mathcal T_j^{(t)} = \tau_{\text{tabu}}$, where $\tau_{\text{tabu}}$ represents the length of tabu period for that variable. This ensures that the variable cannot be flipped again for the next $\tau_{\text{tabu}}$ iterations. For all the other variables that are not flipped, their tabu tenure is decreased by $1$, i.e., $\mathcal T_i^{(t)} = \mathcal T_i^{(t-1)} -1 $, allowing them to gradually become eligible for future flips. It should be noted that $\Delta C_i$ does not need to be recalculated every time, since the flip cost for the rescaled problems changes only locally – only the flip costs of the points connected to $x_i$ are updated. After the whole Tabu search is completed, the algorithm returns a solution along with a list of flip costs. Thereafter, BDSW-QAOA is applied, as discussed above, to solve each sub-QUBO problem using sliding windows over the backbone variables. Overall, an overview of our full framework is depicted in Fig.\ref{fig:overview}. 

Furthermore, our framework offers a flexible mechanism for balancing classical and quantum resources. Specifically, increasing the number of Tabu search iterations can enhance the quality of the initial solution by allowing more thorough exploration of the solution space. In addition, the number of selected backbone variables $k$, as well as the QAOA window size $n_{\text{w}}$, directly determine the extent of quantum resources utilized—namely, the number of required qubits. By tuning these parameters, we can adapt the algorithm to suit different levels of available quantum hardware. When operating on lower-fidelity or smaller-scale NISQ devices, the computational burden can be shifted toward the classical side by performing more extensive Tabu search preprocessing. Conversely, with more advanced quantum hardware, the framework can leverage deeper QAOA circuits and larger quantum windows to reduce classical computation. This dynamic allocation of resources enables scalable optimization under varying hardware constraints and promotes efficient hybrid quantum-classical co-design.

\begin{algorithm}[ht]
\caption{BDSW-QAOA Algorithm}\label{alg:bdsw_qaoa}
\SetAlgoLined
\DontPrintSemicolon
\SetKwInOut{Input}{Input}
\SetKwInOut{Output}{Output}
\SetKwInOut{Initialization}{Initialization} 

\Input{
    $Q$: $N \times N$ QUBO matrix \\
    $N$: number of variables \\
    $\tau_{\text{tabu}}$: tabu tenure \\
    $T$: Tabu search iterations \\
    $k$: backbone variables count \\
    $n$: window size ($n \leq N_{\text{qubit}}$) \\
    QAOA parameters
}
\Output{Optimized solution $\mathbf{x}_{\text{best}}$}
% \BlankLine
% \Initialization{$\mathbf{x}_{\text{current}} \gets$ random binary vector (size $N$) \\
% $\mathcal{T} \gets [0]^N$ (Tabu tenure vector)}
\textbf{Initialization}
$\mathbf{x}_{\text{current}} \gets$ random binary vector (size $N$)\;
$\mathcal{T} \gets [0]^N$ (Tabu tenure vector)\;
Compute $\Delta C_i = (1 - 2x_i)\left(Q_{ii} + \sum_{j \neq i} Q_{ij}x_j\right),\ \forall i$\;

\textbf{Tabu Search Preprocessing:}

\For{$t \gets 1$ \KwTo $T$}{
    Let $S \gets \{i \mid \mathcal{T}_i = 0\}$\tcp*{Non-tabu variables' index set} 
    $j \gets \text{argmax}_{i \in S} |\Delta C_i|$ \tcp*{Max flip cost variable} 
    $x_j \gets 1 - x_j$ \tcp*{Flip} 
    $\mathcal{T}_j \gets \tau_{\text{tabu}}$ \;
    % \ForEach{$i \neq j$}{
    %     \lIf{$\mathcal{T}_i > 0$}{$\mathcal{T}_i \gets \mathcal{T}_i - 1$}
    % }
    \For{each $i \neq j$}{
        \lIf{$\mathcal{T}_i > 0$}{$\mathcal{T}_i \gets \mathcal{T}_i - 1$}
    }
    Update $\Delta C$ for $j$'s neighbors \;
}
$\mathbf{x}_{\text{best}} \gets \mathbf{x}_{\text{current}}$\;
$f_{\text{best}} \gets \mathbf{x}_{\text{best}}^\top Q \mathbf{x}_{\text{best}}$\;

\textbf{Sliding Window QAOA Optimization:}

$\mathcal{B} \gets$ indices of top-$k$ variables by $|\Delta C_i|$ \tcp*{Backbone selection}

\For{$i \gets 0$ \textbf{to} $k - n$}{
    $W \gets \mathcal{B}[i:i+n]$ \tcp*{Sliding window}
    $\text{fixed} \gets \{1,\dots,N\} \setminus W$\;
    Construct $Q_{\text{sub}}$ with: 
    $\quad Q_{\text{sub}}[a,a] \gets Q_{aa} + \sum_{j \in \text{fixed}} Q_{aj}x_j,\ \forall a \in W$\;
    $\mathbf{x}_W^* \gets \text{QAOA}(Q_{\text{sub}})$ \tcp*{Quantum optimization}
    Update $\mathbf{x}_{\text{current}}$ with $\mathbf{x}_W^*$\;
    \If{$\mathbf{x}^\top Q \mathbf{x} < f_{\text{best}}$}{
        $\mathbf{x}_{\text{best}} \gets \mathbf{x}_{\text{current}}$\;
        $f_{\text{best}} \gets \mathbf{x}^\top Q \mathbf{x}$\;
    }
}
\Return{$\mathbf{x}_{\text{best}}$}
\end{algorithm}

\section{Experiments}
Our numerical experiments employ the Max-cut problem as a canonical benchmark for combinatorial optimization due to its well-characterized QUBO formulation and established solution baselines. The evaluation utilizes the G-set and Karloff datasets \cite{yyyeGset,karloff_graphs}, chosen for their controlled complexity progression and demonstrated effectiveness in benchmarking quantum-classical hybrid algorithms \cite{MLQAOA,r32}
% \textcolor{red}{[Cite Some papers]}
. These datasets provide diversity in graph topology and edge weight distribution, enabling a rigorous scalability assessment of our BDSW-QAOA approach.
 \begin{figure}
    \centering
    \includegraphics[width=0.5\textwidth,height=0.24\textheight]{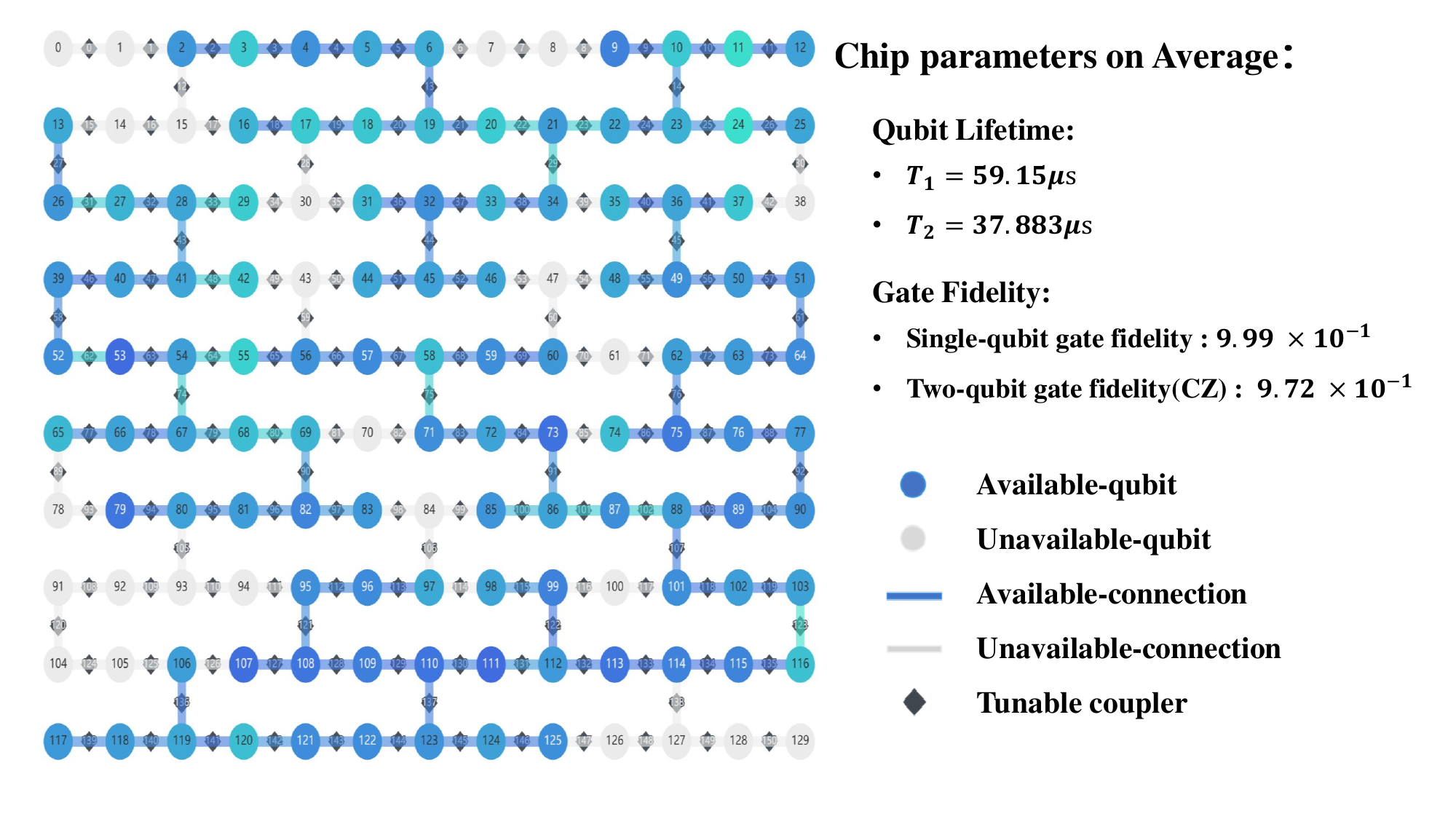}
    \caption{Dongling Quantum Processor Information (Vary over Time).}
    \label{fig:Dongling}
\end{figure}
The QAOA optimization phase is carried out on BAQIS' \textit{Dongling} superconducting quantum processor, accessed through the \textit{Quafu} quantum cloud platform. Figure \ref{fig:Dongling} presents the chip's topological architecture and relevant benchmarking parameters.

For BDSW-QAOA, the window size $n_{\text{w}}$ is set to 15. The number of backbone variables $k$ is determined as a fraction (typically between 0.2 and 0.3) of the total number of variables in the original problem, based on our numerical empirical results. The QAOA circuit depth is fixed at $p = 1$ across all experiments, with a sampling size of 10,240 shots per run. The QAOA circuits are compiled for the Dongling architecture using the compiler strategy adopted in the QCover framework \cite{zhuang2021efficient,xu2024quafu}.
% Tabu search iterations $T$ are controlled by the number of variable flips and the QUBO scale, it is specifically computed as follows: $T = bit\_flips + max(TabuPass\times QUBO\_size)$ , and in the experiments, the value of TabuPass was set to 1700.
% note
% \textcolor{red}{(The detail value?)}. 
 % \cite{zhuang2021efficient,xu2024quafu}.
% And The QAOA's circuit compiler method that maps to dongling is follow the QCover's compiler method \cite{zhuang2021efficient,xu2024quafu}.
We compare our method within two methodological frameworks: classical solvers and quantum hybrid algorithms. The classical solvers comprise dual semidefinite programming (DSDP \cite{DSDP}), graph neural network-based PI-GNN \cite{PIGNN}, break local search (BLS \cite{bls}), and the KHLWG heuristic \cite{KHLWG}. The quantum hybrid algorithms include two variants of multi-level QAOA (MLQAOA)\cite{MLQAOA} and QAOA-in-QAOA\cite{r32}.

\subsection{$G_{set}$ graph}
For our experimental comparison, we selected representative subsets from the $G_\text{set}$ benchmark ($G_1$-$G_5$, $G_{14}$, $G_{15}$, and $G_{22}$) exhibiting distinct topological properties with controlled variations in scale and connectivity. The $G_1$-$G_5$ graphs each contain 800 vertices and 19,716 edges, representing densely connected structures, while $G_{14}$ and $G_{15}$ maintain the same vertex count but with only 4,694 edges to examine sparser connectivity regimes. The larger-scale $G_{22}$ instance (2,000 vertices, 19,990 edges) provides an intermediate edge density case for scalability testing. This systematic selection enables comprehensive evaluation across varying graph topologies, sizes, and edge distributions.

\begin{figure}[h]
    \centering
    \includegraphics[width=0.5\textwidth]{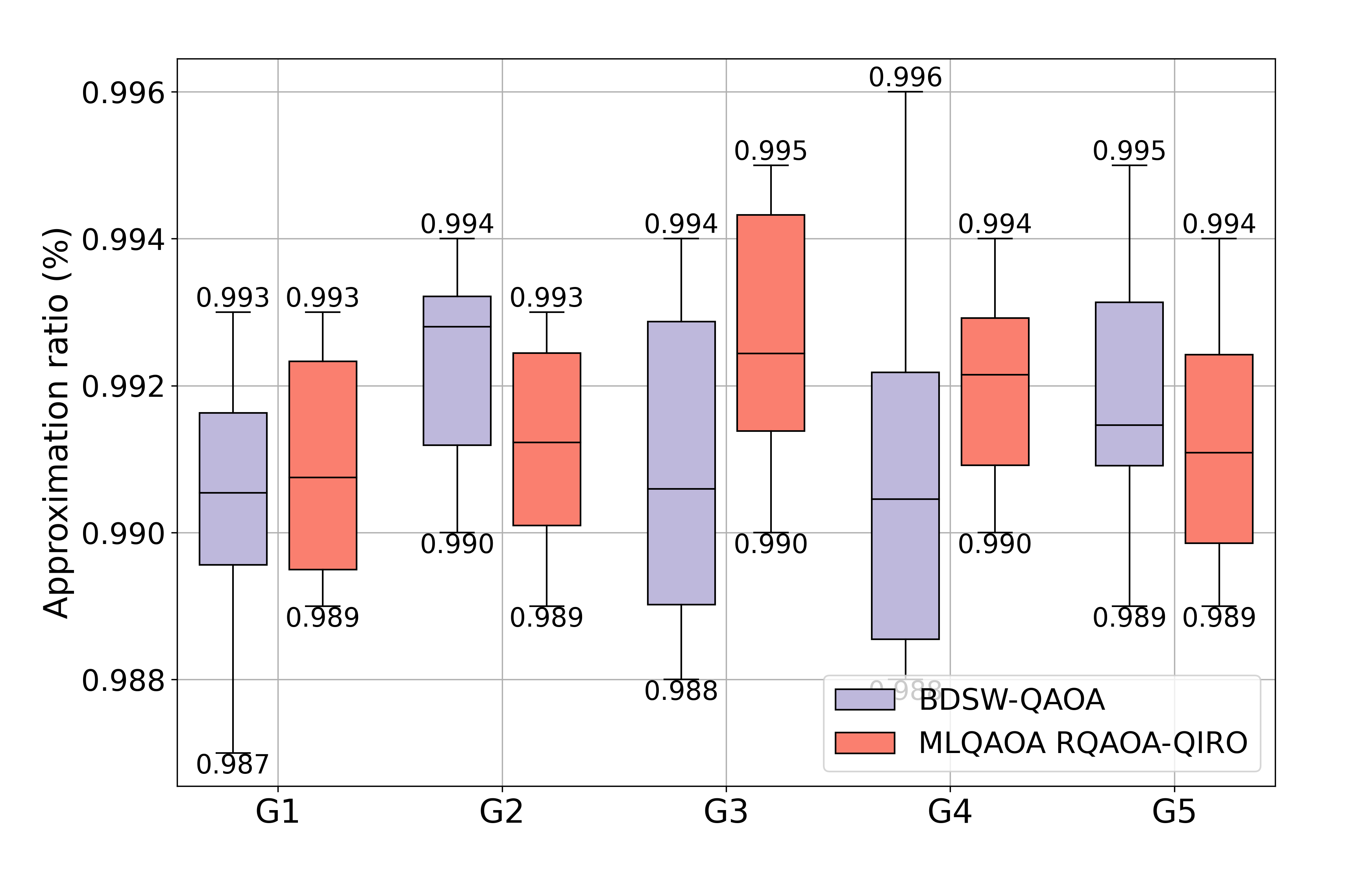}
    \caption{Result of BDSW-QAOA, MLQAOA RQAOA-QIRO in $G_1$ to $G_5$. The calculation of the various quantiles in a box plot is based on the ideal scenario that the data follows a uniform distribution.}
    \label{fig:G_set}
\end{figure}

Table \ref{tab:Gset_table} presents a comparative performance of BDSW-QAOA against two variants of Multi-level QAOA (MLQAOA): the graph learning-based approach and the  Quantum-Informed Recursive Optimization (QIRO) variant inspired by RQAOA \cite{MLQAOA,finvzgar2024quantum,RQAOA}. The evaluation was carried out on graphs $G_1$ through $G_5$ from the $G_{\text{set}}$ benchmark, with each algorithm executed for 20 independent runs to ensure statistical reliability. The results show that BDSW-QAOA consistently achieves competitive approximation ratios (0.987-0.996) in all test cases, outperforming the graph-learning MLQAOA variant (0.970-0.988) in every instance. In particular, the RQAOA-inspired QIRO variant shows performance comparable to BDSW-QAOA (0.989-0.995). The narrow performance variance ($\delta < 0.01$ ) across 20 experimental trials demonstrates algorithmic stability for all tested methods. To provide detailed distributional evidence for these observations, Figure \ref{fig:G_set} compares BDSW-QAOA and MLQAOA QIRO through boxplot visualizations, where boxes represent the interquartile range ($25\% - 75 \%$ percentiles) with median values marked by central lines.

\begin{table}[h]
\centering
\caption{\normalfont Performance comparison between BDSW-QAOA and MLQAOA variants on $G_{\text{set}}$ benchmark graphs. Columns show approximation ratio ranges from 20 independent runs, with final column indicating known optimal MAX-CUT values.}
\label{tab:Gset_table}
\begin{tabular}{c|c|c|c|c}
\hline
$G_{\text{set}}$ & BDSW-QAOA & \multicolumn{2}{c|}{MLQAOA Variants} & Optimal \\
\cline{3-4}
 & & Graph-learning & RQAOA-QIRO & Cut Value \\ 
\hline
$G_1$ & [0.987, 0.993] & [0.976, 0.985] & [0.989, 0.993] & 11624 \\ 
\hline 
$G_2$ & [0.990, 0.994] & [0.978, 0.984] & [0.989, 0.993] & 11620 \\ 
\hline
$G_3$ & [0.994, 0.994] & [0.978, 0.986] & [0.990, 0.995] & 11622 \\ 
\hline
$G_4$ & [0.988, 0.996] & [0.980, 0.988] & [0.990, 0.994] & 11646 \\ 
\hline
$G_5$ & [0.989, 0.995] & [0.970, 0.989] & [0.989, 0.994] & 11631 \\
\hline
\end{tabular}
\end{table}

\begin{table*}[t]
\centering
\caption{$G_{set}$ graphs approximation ratio between different quantum and classical methods}
\label{tab:extenede_Gset}
\small 
\begin{tabular}{c|c|c|c|c|c|c|c|c|c|c}
\hline
\multicolumn{3}{c|}{} & \multicolumn{3}{|c|}{\textbf{Quantum Classical Hybrid}} & \multicolumn{4}{|c|}{\textbf{Classical approach}} & \multicolumn{1}{|c}{\textbf{Optimal}}  \\ \hline
\textbf{$G_{set}$} & $\abs{V}$ & $\abs{E}$ & \begin{tabular}[c]{@{}l@{}}BDSW-QAOA\end{tabular} & \begin{tabular}[c]{@{}l@{}}MLQAOA\\ RQAOA-QIRO\end{tabular} & \begin{tabular}[c]{@{}l@{}}QAOA$^{2}$\end{tabular} & DSDP & PI-GNN  & BLS& KHLWG   \\ \hline
$G_{14}$   & 800   & 4694  & 2988  & 3026 & 2593                                                                        & 2596                                                                        & 2922                      & 3026                      & 3064 & 3064  \\ 
$G_{15}$   & 800   & 4694  & 2953                                                            & 3026                                                      & 2596                                                                        & 2579                                                                        & 2938                      & 2990                      & 3050 & 3050 \\ 
$G_{22}$   & 2000  & 19990 & 13017                                                           & 13174                                                       & 10664                                                                       & 10559                                                                       & 12960                     & 13181                     & 13359  & 13359\\ \hline
\end{tabular}
\end{table*}
% Given that MLQAOA Graph-learning does not yield as many cut scores as MLQAOA RQAOA-QIRO, we decided not to show the results of MLQAOA Graph-learning in Fig \ref{fig:G_set}. Instead, we chose to compare the two methods, BDSW-QAOA and MLQAOA RQAOA-QIRO. This comparison is more effective in highlighting the superior quality of the solution provided by Backbone-Driven QAOA. From the Table \ref{tab:Gset_table} and the Fig \ref{fig:G_set}, we can see that BDSW-QAOA and MLQAOA are comparable in terms of performance, though, when dealing with the $G_{set}$ problem.

\begin{figure}[h]
    \centering
    \includegraphics[width=0.5\textwidth]{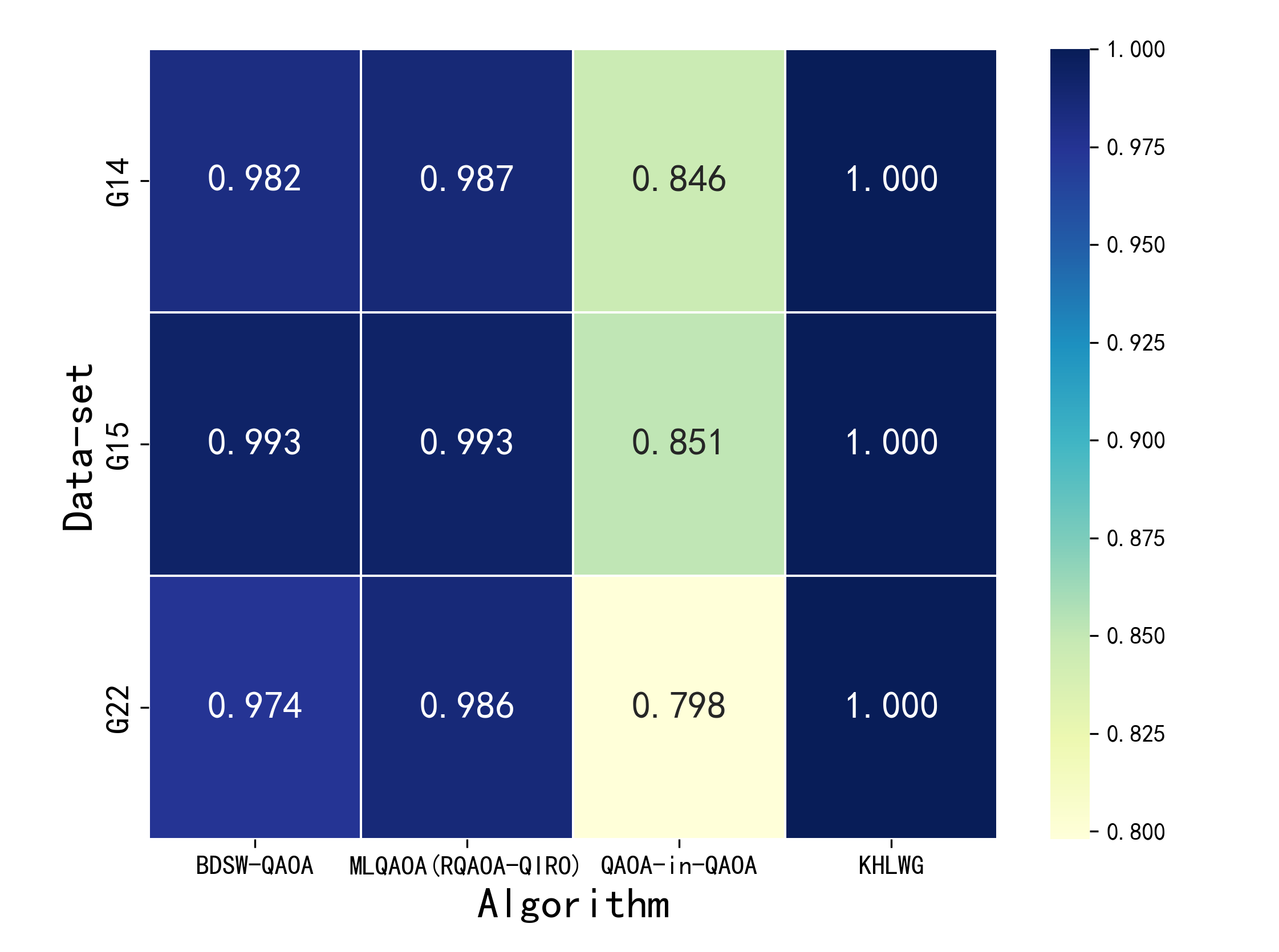}
    \caption{For the comparison data between classical solvers and quantum classical hybrid algorithm on the relatively sparse graphs in $G_{set}$, we only presented the results from the KHLWG in the classical solvers in the chart, due to its superior solution quality.}
    \label{fig:G_14}
\end{figure}

For baseline performance evaluation on the sparse-connectivity graphs $G_{14}$-$G_{15}$ (800 vertices, 4,694 edges) and the large-scale $G_{22}$ instance (2,000 vertices, 19,990 edges). As shown in Table \ref{tab:extenede_Gset}, the KHLWG heuristic demonstrates superior performance among classical approaches. Figure \ref{fig:G_14} specifically contrasts the optimal approximation ratios from 20 experimental trials between KHLWG and quantum-classical hybrid methods to isolate performance frontiers. Experimental comparisons show that BDSW-QAOA and RQAOA-QIRO achieve similar approximation ratios, consistently outperforming QAOA$^2$ by $12\%-15\%$ 
% \textcolor{red}{[check]} 
across all benchmarks. In contrast, the classical KHLWG heuristic attains near-optimal performance with approximation ratios exceeding 0.98
% \textcolor{red}{[Check]} 
— closely approaching the theoretical maximum of 1.0. Notably, the quantum methods’ performance is constrained by the limited fidelity of our shallow $p=1$ QAOA circuits, suggesting deeper circuits or error mitigation could enhance results. 

\begin{figure}[h]
    \centering
    \includegraphics[width=0.5\textwidth]{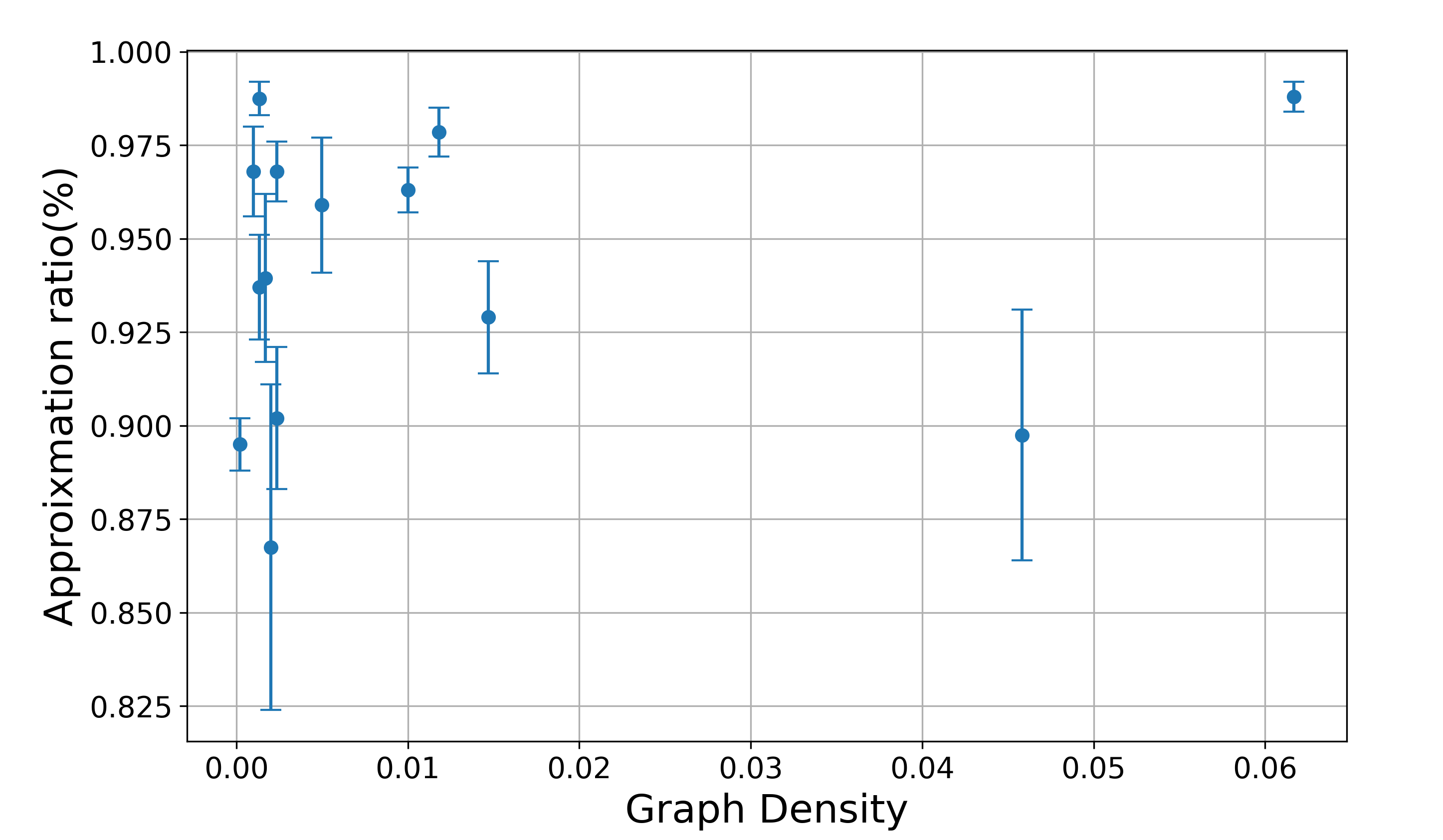}
    \caption{Approximate rate plot for Table IV. The processing power of BDSW-QAOA remains stable with increasing data density.}
    \label{fig:density}
\end{figure}

To further evaluate the performance of BDSW-QAOA, we extended our experiments to cover a broader range of graph densities. Table \ref{g6-g63} presents selected cases from $G_6$ to $G_{63}$, encompassing various levels of sparsity and density, and reports the corresponding range of BDSW-QAOA's approximation ratios. Figure \ref{fig:density} illustrates the relationship between graph density and approximation ratio in these cases, providing a clear view of the algorithm’s performance as density varies. The results indicate that, within our experimental range, BDSW-QAOA remains consistently effective across different graph densities.

\subsection{Karloff graph}
We constructed six test instances from the Karloff graphs, each defined as $K(V, E)$ where $V$ and $E$ denote vertex and edge counts respectively: $K_1(252, 3150)$, $K_2(252, 12600)$, $K_3(924, 16632)$, $K_4(924, 103950)$, $K_5(3432, 84084)$, and $K_6(3432, 756756)$. These instances were selected to benchmark BDSW-QAOA across varying graph scales (vertex counts ranging from 252 to 3432) and topological complexities. Performance comparisons were conducted against MLQAOA and the classical Goemans-Williamson (GW) algorithm, with MLQAOA and GW baseline data extracted from their respective original studies.
% \textcolor{red}{[Cite relevant papers]}. 
The numerical results, summarized in Table \ref{tab:karloff_graph} across 20 independent runs, demonstrate that BDSW-QAOA consistently achieves perfect 1.0 approximation ratios (matching theoretical optima), whereas MLQAOA/RQAOA-QIRO attains near-optimal performance within the range $[0.968, 1.0]$. Notably, BDSW-QAOA maintains its performance even as graph complexity increases by two orders of magnitude in edge density (3,150–756,756 edges) and vertex counts (252–3,432), demonstrating robust scalability of our hybrid algorithms over classical counterparts.

\begin{table}[h]
    \centering
    \caption{Karloff graphs approximation ratio between the Classical and Quantum approach}
    \centering
    \label{tab:karloff_graph}
\begin{tabular}{c|c|c|c|c}
\hline
&\textbf{Classical} & \multicolumn{2}{c}{\textbf{Quantum Classical Hybrid} } & \\ \hline
$K$   & GW \cite{GW}      & \begin{tabular}[c]{@{}l@{}}BDSW-QAOA\end{tabular} & \begin{tabular}[c]{@{}l@{}}MLQAOA\\ RQAOA-QIRO\end{tabular}  &  Optimal\\ \hline
$K_1$ & [0.881 , 0.925] & 1.0                                                             & [0.997 , 1.0]  & 2520                                                             \\ 
$K_2$ & [0.877 , 0.917] & 1.0                                                             & [0.997 , 1.0]  & 12600                                                            \\ 
$K_3$ & [0.879 , 0.913] & 1.0                                                            & [0.994 , 1.0]  & 13860                                                               \\ 
$K_4$ &[0.912 , 0.922]  & 1.0                                                              & [0.995 , 1.0] & 69300                                                               \\
$K_5$ & [0.879 , 0.937] &1.0                                                             &[0.989 , 1.0] & 72072                                                  \\ 
$K_6$ & [0.897 , 0.927] & 1.0                                                             & [0.968 , 1.0]  & 540540                                                         \\ \hline
\end{tabular}
\end{table}
% In Table \ref{tab:karloff_graph}, we observed that for Karloff graphs, Backbone-Driven QAOA can achieve a remarkable approximation ratio of 1. This is primarily due to the characteristics of the Karloff graphs, compared to the $G_{set}$ its frequency of falling into the local optimum is low, after a number of iterations, can reach the optimal approximation rate of 1 and after multiple iterations. This further demonstrates that BDSW-QAOA is superior to MLQAOA in handling sparse graphs.
% However, as the complexity of the graph increases, the advantages of a multilevel optimization framework emerge. resulting in MLQAOA being more capable of handling more complex graphs. But even so, the solution accuracy exhibited by BDSW-QAOA is not much worse than that of MLQAOA.
% However, as the density of the graph increases, the advantages of a multi-level optimization framework gradually come into play \textcolor{red}{[what is multi-level framework?]}. The coarsening and then refining approach is more capable of handling more complex graphs \textcolor{red}{[what is coarening...?]}.

% In other words, BDSW-QAOA maintains a certain level of performance no matter how the scale of the application scenarios changes, and therefore performs well in a variety of different application scenarios \textcolor{red}{[how it can be?]}.

\begin{table}[h]
    \centering
    \caption{Other $G_{set}$ graphs approximation ratio in BDSW-QAOA}
    \centering
    \label{g6-g63}
\begin{tabular}{c|c|c|c|c}
\hline

$G_{set}$& $\abs{V}$ & $\abs{E}$& \begin{tabular}[c]{@{}l@{}}BDSW-QAOA\end{tabular} &  Optimal\\ \hline
$G_6$ & 800 & 19716  & [0.984 , 0.992]&2178                                                             \\ 
$G_{11}$ & 800 & 1600  & [0.864 , 0.931]&564                                                             \\ 
$G_{18}$ & 800 & 4694  & [0.914 , 0.944]&992                                                             \\
$G_{22}$ & 2000 & 19990  & [0.957 , 0.969]&13359                                                             \\
$G_{27}$ & 2000 & 19990  & [0.949 , 0.977]&3341                                                            \\
$G_{32}$  & 2000 & 4000  & [0.824 , 0.911]&1410  \\
$G_{43}$ & 1000 & 9990  & [0.987 , 0.995]&6660   \\                                                           
$G_{48}$ & 3000 & 6000  & [0.983 , 0.992]&6000   \\  
$G_{51}$ & 1000 & 5909  & [0.972 , 0.985]&3850   \\
$G_{55}$ & 5000 & 12498  & [0.921 , 0.966]&10294   \\
$G_{58}$& 5000 & 29570  & [0.960 , 0.976]&19248   \\
$G_{59}$ & 5000 & 29570  & [0.883 , 0.921]&6078   \\
$G_{63}$  & 7000 & 41459 & [0.917 , 0.962]&26963   \\     
$G_{70}$  & 10000 & 9999 & [0.888 , 0.902]&9548   \\      \hline
\end{tabular}
\end{table}

\section{Conclusion}
In this work, we developed BDSW-QAOA, a hybrid quantum-classical algorithm tailored for solving large-scale QUBO problems. Our method combines Tabu-search-based backbone variable selection with a sliding window QAOA framework, enabling dynamic resource allocation between classical pre-processing and quantum optimization. By adjusting the number of backbone variables and the QAOA circuit depth, the algorithm flexibly leverages the limited qubit capacity and fidelity of current NISQ devices, while maintaining high solution quality.

Experimental evaluations on diverse benchmark graphs—from the G-set to Karloff instances—demonstrate that BDSW-QAOA consistently achieves near-optimal approximation ratios across various graph densities. This robustness suggests that our resource allocation strategy effectively adapts to the inherent variability of real-world optimization problems. Even when graph density increases, BDSW-QAOA’s performance remains stable, reinforcing the notion that dynamic interplay between quantum and classical components is key for overcoming the constraints of shallow quantum circuits. Moreover, our results highlight the potential benefits of enhancing either the classical or the quantum part of the hybrid algorithm. For instance, increasing the precision of the classical Tabu search can further improve backbone selection, while an upgraded quantum hardware platform would allow the use of a larger number of backbone variables and deeper QAOA circuits. These complementary improvements could work synergistically to push the performance closer to theoretical optima.

Overall, this study provides a practical framework for quantum-classical co-optimization in the NISQ era. By demonstrating the effectiveness and scalability of BDSW-QAOA, our research lays a solid foundation for future work aimed at integrating advanced error mitigation and deeper quantum circuits, ultimately moving toward achieving a quantum advantage in complex combinatorial optimization tasks.

\section*{Acknowledgment}
This work is supported by the National Natural Science Foundation of China (Grant No. 92365111 and 62101600), the Beijing Natural Science Foundation (Grant No. Z220002), and the Innovation Program for Quantum Science and Technology (Grant No. 2021ZD0302400). Additional support is provided by the State Key Lab of Processors at the Institute of Computing Technology, Chinese Academy of Sciences (Grant No. CLQ202404), and the Science Foundation of the China University of Petroleum, Beijing (Grant No. 2462021YJRC008). The authors gratefully acknowledge all funding agencies for their contributions to this research.

% This research was supported by the National Nature Science Foundation of China (Grants No.62101600), State Key Lab of Processors,Institute of Computing Technology,CAS under Grant No.CLQ202404,the Science Foundation of the China University of Petroleum,Beijing (Grant No. 2462021YJRC008).
% \bibliographystyle{unsrt}
\bibliographystyle{IEEEtran}
\bibliography{bibliography}

\begin{thebibliography}{10}
\providecommand{\url}[1]{#1}
\csname url@samestyle\endcsname
\providecommand{\newblock}{\relax}
\providecommand{\bibinfo}[2]{#2}
\providecommand{\BIBentrySTDinterwordspacing}{\spaceskip=0pt\relax}
\providecommand{\BIBentryALTinterwordstretchfactor}{4}
\providecommand{\BIBentryALTinterwordspacing}{\spaceskip=\fontdimen2\font plus
\BIBentryALTinterwordstretchfactor\fontdimen3\font minus
  \fontdimen4\font\relax}
\providecommand{\BIBforeignlanguage}[2]{{%
\expandafter\ifx\csname l@#1\endcsname\relax
\typeout{** WARNING: IEEEtran.bst: No hyphenation pattern has been}%
\typeout{** loaded for the language `#1'. Using the pattern for}%
\typeout{** the default language instead.}%
\else
\language=\csname l@#1\endcsname
\fi
#2}}
\providecommand{\BIBdecl}{\relax}
\BIBdecl

\bibitem{markowitz1952portfolio}
\BIBentryALTinterwordspacing
H.~M. Markowitz, \emph{Portfolio Selection: Efficient Diversification of
  Investments}.\hskip 1em plus 0.5em minus 0.4em\relax Yale University Press,
  1959. [Online]. Available: \url{http://www.jstor.org/stable/j.ctt1bh4c8h}
\BIBentrySTDinterwordspacing

\bibitem{abdalkareem2021healthcare}
\BIBentryALTinterwordspacing
Z.~A. Abdalkareem, A.~Amir, M.~A. Al-Betar, P.~Ekhan, and A.~I. Hammouri,
  ``Healthcare scheduling in optimization context: a review,'' \emph{Health and
  Technology}, vol.~11, pp. 445--469, 2021. [Online]. Available:
  \url{https://link.springer.com/article/10.1007/s12553-021-00547-5}
\BIBentrySTDinterwordspacing

\bibitem{rbioinformatics}
\BIBentryALTinterwordspacing
J.~Handl, D.~B. Kell, and J.~Knowles, ``Multiobjective optimization in
  bioinformatics and computational biology,'' \emph{IEEE/ACM Transactions on
  computational biology and bioinformatics}, vol.~4, no.~2, pp. 279--292, 2007.
  [Online]. Available: \url{https://ieeexplore.ieee.org/document/4196538}
\BIBentrySTDinterwordspacing

\bibitem{r51}
\BIBentryALTinterwordspacing
A.~Sbihi and R.~W. Eglese, ``Combinatorial optimization and green logistics,''
  \emph{Annals of Operations Research}, vol. 175, pp. 159--175, 2010. [Online].
  Available: \url{https://doi.org/10.1007/s10479-009-0651-z}
\BIBentrySTDinterwordspacing

\bibitem{np}
\BIBentryALTinterwordspacing
D.~S. Hochba, ``Approximation algorithms for np-hard problems,'' \emph{ACM
  Sigact News}, vol.~28, no.~2, pp. 40--52, 1997. [Online]. Available:
  \url{https://doi.org/10.1145/261342.571216}
\BIBentrySTDinterwordspacing

\bibitem{hartmanis1982computers}
J.~Hartmanis, ``Computers and intractability: a guide to the theory of
  np-completeness (michael r. garey and david s. johnson),'' \emph{Siam
  Review}, vol.~24, no.~1, p.~90, 1982.

\bibitem{ACO}
\BIBentryALTinterwordspacing
S.~Fenet and C.~Solnon, ``Searching for maximum cliques with ant colony
  optimization,'' in \emph{Applications of Evolutionary Computing: EvoWorkshops
  2003: EvoBIO, EvoCOP, EvoIASP, EvoMUSART, EvoROB, and EvoSTIM Essex, UK,
  April 14--16, 2003 Proceedings}.\hskip 1em plus 0.5em minus 0.4em\relax
  Springer, 2003, pp. 236--245. [Online]. Available:
  \url{https://link.springer.com/chapter/10.1007/3-540-36605-9_22}
\BIBentrySTDinterwordspacing

\bibitem{GA}
\BIBentryALTinterwordspacing
G.~Zhang, L.~Zhang, X.~Song, Y.~Wang, and C.~Zhou, ``A variable neighborhood
  search based genetic algorithm for flexible job shop scheduling problem,''
  \emph{Cluster Computing}, vol.~22, pp. 11\,561--11\,572, 2019. [Online].
  Available: \url{https://doi.org/10.1007/s10586-017-1420-4}
\BIBentrySTDinterwordspacing

\bibitem{r49}
\BIBentryALTinterwordspacing
Y.~He, Y.~Qiu, and G.~Liu, ``A tabu search approach with double tabu-list for
  multidimensional knapsack problems,'' \emph{IJCSNS International Journal of
  Computer Science and Network Security}, vol.~6, no.~5A, pp. 87--92, 2006.
  [Online]. Available:
  \url{https://citeseerx.ist.psu.edu/document?repid=rep1&type=pdf&doi=4899fa7560b77f231a0b6e090aec406874d68f85}
\BIBentrySTDinterwordspacing

\bibitem{shor1994algorithms}
P.~W. Shor, ``Algorithms for quantum computation: discrete logarithms and
  factoring,'' in \emph{Proceedings 35th annual symposium on foundations of
  computer science}.\hskip 1em plus 0.5em minus 0.4em\relax Ieee, 1994, pp.
  124--134.

\bibitem{arute2019quantum}
\BIBentryALTinterwordspacing
F.~Arute, K.~Arya, R.~Babbush, D.~Bacon, J.~C. Bardin, R.~Barends, R.~Biswas,
  S.~Boixo, F.~G. Brandao, D.~A. Buell \emph{et~al.}, ``Quantum supremacy using
  a programmable superconducting processor,'' \emph{Nature}, vol. 574, no.
  7779, pp. 505--510, 2019. [Online]. Available:
  \url{https://www.nature.com/articles/s41586-019-1666-5}
\BIBentrySTDinterwordspacing

\bibitem{googlewillow}
\BIBentryALTinterwordspacing
G.~Q. AI and Collaborators, ``Quantum error correction below the surface code
  threshold,'' \emph{Nature}, vol. 638, pp. 920--926, 2025. [Online].
  Available: \url{https://www.nature.com/articles/s41586-024-08449-y}
\BIBentrySTDinterwordspacing

\bibitem{chitambar2019quantum}
\BIBentryALTinterwordspacing
E.~Chitambar and G.~Gour, ``Quantum resource theories,'' \emph{Reviews of
  modern physics}, vol.~91, no.~2, p. 025001, 2019. [Online]. Available:
  \url{https://journals.aps.org/rmp/abstract/10.1103/RevModPhys.91.025001}
\BIBentrySTDinterwordspacing

\bibitem{r54}
\BIBentryALTinterwordspacing
R.~Au-Yeung, N.~Chancellor, and P.~Halffmann, ``Np-hard but no longer hard to
  solve? using quantum computing to tackle optimization problems,''
  \emph{Frontiers in Quantum Science and Technology}, vol.~2, p. 1128576, 2023.
  [Online]. Available: \url{https://doi.org/10.3389/frqst.2023.1128576}
\BIBentrySTDinterwordspacing

\bibitem{brooks2019beyond}
M.~Brooks, ``Beyond quantum supremacy: the hunt for useful quantum computers,''
  \emph{Nature}, vol. 574, no. 7776, pp. 19--22, 2019.

\bibitem{r11}
\BIBentryALTinterwordspacing
E.~Farhi, J.~Goldstone, and S.~Gutmann, ``\BIBforeignlanguage{en-US}{A quantum
  approximate optimization algorithm},''
  \emph{\BIBforeignlanguage{en-US}{arXiv: Quantum Physics,arXiv: Quantum
  Physics}}, Nov 2014. [Online]. Available:
  \url{https://doi.org/10.48550/arXiv.1411.4028}
\BIBentrySTDinterwordspacing

\bibitem{peruzzo2014variational}
A.~Peruzzo, J.~McClean, P.~Shadbolt, M.-H. Yung, X.-Q. Zhou, P.~J. Love,
  A.~Aspuru-Guzik, and J.~L. O’brien, ``A variational eigenvalue solver on a
  photonic quantum processor,'' \emph{Nature communications}, vol.~5, no.~1, p.
  4213, 2014.

\bibitem{biamonte2017quantum}
J.~Biamonte, P.~Wittek, N.~Pancotti, P.~Rebentrost, N.~Wiebe, and S.~Lloyd,
  ``Quantum machine learning,'' \emph{Nature}, vol. 549, no. 7671, pp.
  195--202, 2017.

\bibitem{warm-start}
\BIBentryALTinterwordspacing
D.~J. Egger, J.~Mareček, and S.~Woerner,
  ``\BIBforeignlanguage{en-US}{Warm-starting quantum optimization},''
  \emph{\BIBforeignlanguage{en-US}{Quantum}}, p. 479, Jun 2021. [Online].
  Available: \url{http://dx.doi.org/10.22331/q-2021-06-17-479}
\BIBentrySTDinterwordspacing

\bibitem{gnn-qaoa}
\BIBentryALTinterwordspacing
N.~Jain, B.~Coyle, E.~Kashefi, and N.~Kumar, ``Graph neural network
  initialisation of quantum approximate optimisation,'' \emph{Quantum}, vol.~6,
  p. 861, 2022. [Online]. Available:
  \url{https://doi.org/10.22331/q-2022-11-17-861}
\BIBentrySTDinterwordspacing

\bibitem{qa-qaoa}
\BIBentryALTinterwordspacing
S.~H. Sack and M.~Serbyn, ``Quantum annealing initialization of the quantum
  approximate optimization algorithm,'' \emph{quantum}, vol.~5, p. 491, 2021.
  [Online]. Available: \url{https://doi.org/10.22331/q-2021-07-01-491}
\BIBentrySTDinterwordspacing

\bibitem{alignment}
\BIBentryALTinterwordspacing
Z.~He, R.~Shaydulin, S.~Chakrabarti, D.~Herman, C.~Li, Y.~Sun, and M.~Pistoia,
  ``Alignment between initial state and mixer improves qaoa performance for
  constrained optimization,'' \emph{npj Quantum Information}, vol.~9, no.~1, p.
  121, 2023. [Online]. Available:
  \url{https://doi.org/10.1038/s41534-023-00787-5}
\BIBentrySTDinterwordspacing

\bibitem{r25}
\BIBentryALTinterwordspacing
L.~Zhou, S.-T. Wang, S.~Choi, H.~Pichler, and M.~D. Lukin,
  ``\BIBforeignlanguage{en-US}{Quantum approximate optimization algorithm:
  Performance, mechanism, and implementation on near-term devices},''
  \emph{\BIBforeignlanguage{en-US}{Physical Review X}}, Jun 2020. [Online].
  Available: \url{http://dx.doi.org/10.1103/physrevx.10.021067}
\BIBentrySTDinterwordspacing

\bibitem{greedy-qaoa}
\BIBentryALTinterwordspacing
S.~H. Sack, R.~A. Medina, R.~Kueng, and M.~Serbyn, ``Recursive greedy
  initialization of the quantum approximate optimization algorithm with
  guaranteed improvement,'' \emph{Physical Review A}, vol. 107, no.~6, p.
  062404, 2023. [Online]. Available:
  \url{https://doi.org/10.1103/PhysRevA.107.062404}
\BIBentrySTDinterwordspacing

\bibitem{depth-qaoa}
\BIBentryALTinterwordspacing
X.~Lee, N.~Xie, D.~Cai, Y.~Saito, and N.~Asai, ``A depth-progressive
  initialization strategy for quantum approximate optimization algorithm,''
  \emph{Mathematics}, vol.~11, no.~9, p. 2176, 2023. [Online]. Available:
  \url{https://doi.org/10.3390/math11092176}
\BIBentrySTDinterwordspacing

\bibitem{r31}
\BIBentryALTinterwordspacing
H.~Ushijima-Mwesigwa, R.~Shaydulin, C.~F.~A. Negre, S.~M. Mniszewski,
  Y.~Alexeev, and I.~Safro, ``\BIBforeignlanguage{en-US}{Multilevel
  combinatorial optimization across quantum architectures},''
  \emph{\BIBforeignlanguage{en-US}{ACM Transactions on Quantum Computing}}, p.
  1–29, Mar 2021. [Online]. Available:
  \url{http://dx.doi.org/10.1145/3425607}
\BIBentrySTDinterwordspacing

\bibitem{r32}
\BIBentryALTinterwordspacing
Z.~Zhou, Y.~Du, X.~Tian, and D.~Tao, ``\BIBforeignlanguage{en-US}{Qaoa-in-qaoa:
  solving large-scale maxcut problems on small quantum machines},'' May 2022.
  [Online]. Available: \url{https://doi.org/10.1103/PhysRevApplied.19.024027}
\BIBentrySTDinterwordspacing

\bibitem{boost2017partitioning}
M.~Boost, S.~Reinhardt, and A.~Roy, ``Partitioning optimization problems for
  hybrid classical/quantum execution,'' \emph{D-Wave Syst., D-Wave Technical
  Report Series, Burnaby, CO, Canada, Tech. Rep}, 2017.

\bibitem{r2}
\BIBentryALTinterwordspacing
F.~Glover, G.~Kochenberger, and Y.~Du, ``A tutorial on formulating and using
  qubo models,'' 2019. [Online]. Available:
  \url{https://doi.org/10.48550/arXiv.1811.11538}
\BIBentrySTDinterwordspacing

\bibitem{QUBO}
\BIBentryALTinterwordspacing
F.~W. Glover and G.~A. Kochenberger, ``A tutorial on formulating {QUBO}
  models,'' \emph{CoRR}, vol. abs/1811.11538, 2018. [Online]. Available:
  \url{http://arxiv.org/abs/1811.11538}
\BIBentrySTDinterwordspacing

\bibitem{r19}
\BIBentryALTinterwordspacing
K.~Jun, ``Qubo formulations for numerical quantum computing,'' 2022. [Online].
  Available: \url{https://arxiv.org/abs/2106.10819}
\BIBentrySTDinterwordspacing

\bibitem{d-wave}
\BIBentryALTinterwordspacing
H.~Ushijima-Mwesigwa, C.~F. Negre, and S.~M. Mniszewski, ``Graph partitioning
  using quantum annealing on the d-wave system,'' in \emph{Proceedings of the
  Second International Workshop on Post Moores Era Supercomputing}, 2017, pp.
  22--29. [Online]. Available: \url{https://doi.org/10.1145/3149526.3149531}
\BIBentrySTDinterwordspacing

\bibitem{r20}
\BIBentryALTinterwordspacing
G.~G. Guerreschi, ``Solving quadratic unconstrained binary optimization with
  divide-and-conquer and quantum algorithms,'' 2021. [Online]. Available:
  \url{https://arxiv.org/abs/2101.07813}
\BIBentrySTDinterwordspacing

\bibitem{glover1986future}
\BIBentryALTinterwordspacing
F.~Glover, ``Future paths for integer programming and links to artificial
  intelligence,'' \emph{Computers \& operations research}, vol.~13, no.~5, pp.
  533--549, 1986. [Online]. Available:
  \url{https://doi.org/10.1016/0305-0548(86)90048-1}
\BIBentrySTDinterwordspacing

\bibitem{glover1997tabu}
\BIBentryALTinterwordspacing
------, ``Tabu search and adaptive memory programming—advances, applications
  and challenges,'' \emph{Interfaces in computer science and operations
  research: Advances in metaheuristics, optimization, and stochastic modeling
  technologies}, pp. 1--75, 1997. [Online]. Available:
  \url{https://doi.org/10.1007/978-1-4615-4102-8_1}
\BIBentrySTDinterwordspacing

\bibitem{r2005tabu}
\BIBentryALTinterwordspacing
M.~Gendreau and J.-Y. Potvin, ``Tabu search,'' \emph{Search methodologies:
  introductory tutorials in optimization and decision support techniques}, pp.
  165--186, 2005. [Online]. Available:
  \url{https://doi.org/10.1007/0-387-28356-0_6}
\BIBentrySTDinterwordspacing

\bibitem{rtabututorial}
\BIBentryALTinterwordspacing
F.~Glover, ``Tabu search: A tutorial,'' \emph{Interfaces}, vol.~20, no.~4, pp.
  74--94, 1990. [Online]. Available: \url{https://doi.org/10.1287/inte.20.4.74}
\BIBentrySTDinterwordspacing

\bibitem{blekos2024review}
\BIBentryALTinterwordspacing
K.~Blekos, D.~Brand, A.~Ceschini, C.-H. Chou, R.-H. Li, K.~Pandya, and
  A.~Summer, ``A review on quantum approximate optimization algorithm and its
  variants,'' \emph{Physics Reports}, vol. 1068, pp. 1--66, 2024. [Online].
  Available: \url{https://doi.org/10.1016/j.physrep.2024.03.002}
\BIBentrySTDinterwordspacing

\bibitem{lotshaw2021empirical}
\BIBentryALTinterwordspacing
P.~C. Lotshaw, T.~S. Humble, R.~Herrman, J.~Ostrowski, and G.~Siopsis,
  ``Empirical performance bounds for quantum approximate optimization,''
  \emph{Quantum Information Processing}, vol.~20, no.~12, p. 403, 2021.
  [Online]. Available: \url{https://doi.org/10.1007/s11128-021-03342-3}
\BIBentrySTDinterwordspacing

\bibitem{benchasattabuse2023lower}
\BIBentryALTinterwordspacing
N.~Benchasattabuse, A.~Bärtschi, L.~P. García-Pintos, J.~Golden, N.~Lemons,
  and S.~Eidenbenz, ``Lower bounds on number of qaoa rounds required for
  guaranteed approximation ratios,'' 2023. [Online]. Available:
  \url{https://arxiv.org/abs/2308.15442}
\BIBentrySTDinterwordspacing

\bibitem{r28}
\BIBentryALTinterwordspacing
F.~Glover, \emph{\BIBforeignlanguage{en-US}{Adaptive Memory Projection Methods
  for Integer Programming}}, Aug 2005, p. 425–440. [Online]. Available:
  \url{http://dx.doi.org/10.1007/0-387-23667-8_19}
\BIBentrySTDinterwordspacing

\bibitem{r29}
\BIBentryALTinterwordspacing
------, ``\BIBforeignlanguage{en-US}{Heuristics for integer programming using
  surrogate constraints},'' \emph{\BIBforeignlanguage{en-US}{Decision
  Sciences}}, p. 156–166, Jan 1977. [Online]. Available:
  \url{http://dx.doi.org/10.1111/j.1540-5915.1977.tb01074.x}
\BIBentrySTDinterwordspacing

\bibitem{glover2005adaptive}
\BIBentryALTinterwordspacing
------, ``Adaptive memory projection methods for integer programming,''
  \emph{Metaheuristic optimization via memory and evolution: Tabu search and
  scatter search}, pp. 425--440, 2005. [Online]. Available:
  \url{https://link.springer.com/chapter/10.1007/0-387-23667-8_19}
\BIBentrySTDinterwordspacing

\bibitem{yyyeGset}
Y.~Ye, ``{Gset} - a suite-style benchmark for graph processing systems,''
  \url{https://web.stanford.edu/~yyye/yyye/Gset/}, 2003.

\bibitem{karloff_graphs}
\BIBentryALTinterwordspacing
H.~Karloff, ``{How good is the Goemans-Williamson MAX CUT algorithm?}'' in
  \emph{Proceedings of the Twenty-Eighth Annual ACM Symposium on Theory of
  Computing}, ser. STOC '96, 1996, p. 427–434. [Online]. Available:
  \url{https://dl.acm.org/doi/pdf/10.1145/237814.237990}
\BIBentrySTDinterwordspacing

\bibitem{MLQAOA}
\BIBentryALTinterwordspacing
B.~Bach, J.~Falla, and I.~Safro, ``Mlqaoa: Graph learning accelerated hybrid
  quantum-classical multilevel qaoa,'' 2024. [Online]. Available:
  \url{https://ieeexplore.ieee.org/abstract/document/10821278}
\BIBentrySTDinterwordspacing

\bibitem{zhuang2021efficient}
W.-F. Zhuang, Y.-N. Pu, H.-Z. Xu, X.~Chai, Y.~Gu, Y.~Ma, S.~Qamar, C.~Qian,
  P.~Qian, X.~Xiao \emph{et~al.}, ``Efficient classical computation of quantum
  mean values for shallow qaoa circuits,'' \emph{arXiv preprint
  arXiv:2112.11151}, 2021.

\bibitem{xu2024quafu}
\BIBentryALTinterwordspacing
H.-Z. Xu, W.-F. Zhuang, Z.-A. Wang, K.-X. Huang, Y.-H. Shi, W.-G. Ma, T.-M. Li,
  C.-T. Chen, K.~Xu, Y.-L. Feng \emph{et~al.}, ``Quafu-qcover: Explore
  combinatorial optimization problems on cloud-based quantum computers,''
  \emph{Chinese Physics B}, vol.~33, no.~5, p. 050302, 2024. [Online].
  Available: \url{https://iopscience.iop.org/article/10.1088/1674-1056/ad18ab}
\BIBentrySTDinterwordspacing

\bibitem{DSDP}
\BIBentryALTinterwordspacing
S.~J. Benson, Y.~Ye, and X.~Zhang, ``Solving large-scale sparse semidefinite
  programs for combinatorial optimization,'' \emph{SIAM Journal on
  Optimization}, vol.~10, no.~2, pp. 443--461, 2000. [Online]. Available:
  \url{https://doi.org/10.1137/S1052623497328008}
\BIBentrySTDinterwordspacing

\bibitem{PIGNN}
\BIBentryALTinterwordspacing
M.~J.~A. Schuetz, J.~K. Brubaker, and H.~G. Katzgraber, ``Combinatorial
  optimization with physics-inspired graph neural networks,'' \emph{Nature
  Machine Intelligence}, vol.~4, no.~4, p. 367–377, Apr. 2022. [Online].
  Available: \url{http://dx.doi.org/10.1038/s42256-022-00468-6}
\BIBentrySTDinterwordspacing

\bibitem{bls}
\BIBentryALTinterwordspacing
U.~Benlic and J.-K. Hao, ``Breakout local search for the max-cutproblem,''
  \emph{Engineering Applications of Artificial Intelligence}, vol.~26, no.~3,
  pp. 1162--1173, 2013. [Online]. Available:
  \url{https://www.sciencedirect.com/science/article/pii/S0952197612002175}
\BIBentrySTDinterwordspacing

\bibitem{KHLWG}
\BIBentryALTinterwordspacing
G.~Kochenberger, J.-K. Hao, Z.~Lü, H.~Wang, and F.~Glover, ``Solving large
  scale max cut problems via tabu search,'' \emph{Journal of Heuristics},
  vol.~19, 08 2013. [Online]. Available:
  \url{https://doi.org/10.1007/s10732-011-9189-8}
\BIBentrySTDinterwordspacing

\bibitem{finvzgar2024quantum}
\BIBentryALTinterwordspacing
J.~R. Fin{\v{z}}gar, A.~Kerschbaumer, M.~J. Schuetz, C.~B. Mendl, and H.~G.
  Katzgraber, ``Quantum-informed recursive optimization algorithms,'' \emph{PRX
  Quantum}, vol.~5, no.~2, p. 020327, 2024. [Online]. Available:
  \url{https://doi.org/10.1103/PRXQuantum.5.020327}
\BIBentrySTDinterwordspacing

\bibitem{RQAOA}
\BIBentryALTinterwordspacing
S.~Bravyi, A.~Kliesch, R.~Koenig, and E.~Tang, ``Obstacles to variational
  quantum optimization from symmetry protection,'' \emph{Phys. Rev. Lett.},
  vol. 125, p. 260505, Dec 2020. [Online]. Available:
  \url{https://link.aps.org/doi/10.1103/PhysRevLett.125.260505}
\BIBentrySTDinterwordspacing

\bibitem{GW}
\BIBentryALTinterwordspacing
M.~X. Goemans and D.~P. Williamson, ``Improved approximation algorithms for
  maximum cut and satisfiability problems using semidefinite programming,''
  \emph{J. ACM}, vol.~42, no.~6, p. 1115–1145, nov 1995. [Online]. Available:
  \url{https://doi.org/10.1145/227683.227684}
\BIBentrySTDinterwordspacing

\end{thebibliography}
\end{document}